\documentclass{Interspeech2024}
\usepackage{amsmath,graphicx,url}
\usepackage{subcaption}

\interspeechcameraready

\title{LSTMSE-Net: Long Short Term Speech Enhancement Network for Audio-visual Speech Enhancement}
\name[affiliation={1}]{Arnav}{Jain$^*$}
\name[affiliation={1}]{Jasmer}{Singh Sanjotra$^*$}
\name[affiliation={1}]{Harshvardhan}{Choudhary}
\name[affiliation={1}]{Krish}{Agrawal}
\name[affiliation={1}]{Rupal}{Shah}
\name[affiliation={1}]{Rohan}{Jha}
\name[affiliation={1}]{M.}{Sajid}
\name[affiliation={2}]{Amir}{Hussain}
\name[affiliation={1}]{M.}{Tanveer}

\address{
  $^1$Indian Institute of Technology Indore, Simrol, Indore, 453552, India\\
  $^2$School of Computing, Edinburgh Napier University, EH11 4BN, Edinburgh, United Kingdom}
\email{A.Hussain@napier.ac.uk, mtanveer@iiti.ac.in}


\begin{document}
\maketitle

\begin{abstract}
In this paper, we propose long short term memory speech enhancement network (LSTMSE-Net), an audio-visual speech enhancement (AVSE) method. This innovative method leverages the complementary nature of visual and audio information to boost the quality of speech signals. Visual features are extracted with VisualFeatNet (VFN), and audio features are processed through an encoder and decoder. The system scales and concatenates visual and audio features, then processes them through a separator network for optimized speech enhancement. The architecture highlights advancements in leveraging multi-modal data and interpolation techniques for robust AVSE challenge systems. The performance of LSTMSE-Net surpasses that of the baseline model from the COG-MHEAR AVSE Challenge 2024 by a margin of 0.06 in scale-invariant signal-to-distortion ratio (SISDR), $0.03$ in short-time objective intelligibility (STOI), and $1.32$ in perceptual evaluation of speech quality (PESQ). The source code of the proposed LSTMSE-Net is available at \url{https://github.com/mtanveer1/AVSEC-3-Challenge}.
\end{abstract}
%%%%%%%%%%%%%%%%
\renewcommand{\thefootnote}{\fnsymbol{footnote}} % This changes the numbering style to symbols
\footnotetext[1]{These authors contributed equally to this work.}
%%%%%%%%%%%%%%%%%%%%%%%
\section{Introduction}
Speech is key to how humans interact. Speech clarity and quality are critical for domains like video conferencing, telecommunications, voice assistants, hearing aids, etc. However, maintaining high-quality speech in adverse acoustic conditions—such as environments with background noise, reverberation, or poor audio quality—remains a significant challenge. Speech enhancement (SE) has become a pivotal area of study and development to solve these problems and enhance speech quality and intelligibility \cite{loizou2007speech}. Deep learning approaches have been the driving force behind recent advances in SE. While deep learning-based SE techniques \cite{lu2013speech,6853860} have shown exceptional success by focusing mainly on audio signals, it is crucial to understand that adding visual information can greatly improve SE systems' performance in adverse sound conditions. \cite{8323326,9156852,10.1109/TASLP.2021.3066303}. For comprehensive insights into speech signal processing tasks using ensemble deep learning methods, readers are referred to \cite{tanveer2023ensemble}.

Time-frequency (TF) domain methods and time-domain methods are two general categories into which audio-only SE methods can be divided, depending on the type of input. Classical TF domain techniques often rely on amplitude spectrum features; however, studies shows that their effectiveness may be constrained if phase information is not taken into account \cite{6853860}. Some methods that make use of complex-valued features have been introduced to get around this restriction, including complex spectral mapping (CSM) \cite{8682834} and complex ratio masking (CRM) \cite{williamson2017time}. Real-valued neural networks are used in the implementation of many CRM and CSM techniques, whereas neural networks using complex values are used in other cases to handle complex input. Notable examples of complex-valued neural networks for SE tasks include deep complex convolution recurrent network (DCCRN) \cite{hu20g_interspeech} and deep complex U-NET (DCUNET) \cite{choi2018phaseaware}. In this study, we employ time-domain-based methods as well as real-valued neural networks to show their effectiveness in SE tasks.

The primary idea in the wake of audio-visual speech enhancement (AVSE) is to augment an audio-only SE system with visual input as supplemental data with the goal of improving SE performance. The advantage of using visual input to enhance SE system performance has been demonstrated in a number of earlier research \cite{8323326,DBLP:journals/corr/abs-1804-04121,chuang2022improved}. Most preceding AVSE methods focused on processing audio in the TF domain \cite{10193049,10.1109/TASLP.2021.3066303}, however some research have explored time domain methods for audio-visual speech separation tasks \cite{9746866}. Additionally, techniques such as self supervised learning (SSL) embedding are used to boost AVSE performance. Richard et al. \cite{lai2023audiovisual} presented the SSL-AVSE technique, which combines auditory and visual cues. These combined audio-visual features are then analyzed by a Transformer-based SSL AV-HuBERT model to extract characteristics, which are then controlled by a BLSTM-based SE model. However, these models are too large to be scalable or deployable in real-life scenarios. Therefore, we focused on developing a smaller, simpler and a scalable model that maintains performance comparable to these larger models.

% Our research builds upon the baseline model provided by the AVSE Challenge, which employs a combination of visual feature neural networks, UNETs\cite{ronneberger2015u}, and Temporal Convolutional Networks (TCNs)\cite{lea2017temporal} to perform audio-visual speech separation. To further improve the performance of this baseline model, we incorporated Long Short-Term Memory (LSTM) networks. This enhancement resulted in a significant improvement in key performance metrics, demonstrating the efficacy of our approach.
% This research contributes to the advancement of multimodal speech processing by demonstrating the benefits of integrating visual information with audio signals. Our enhanced model showcases the potential of using advanced neural network architectures to improve speech intelligibility and quality in noisy environments, paving the way for more robust and effective speech-based technologies.

In this paper, we propose long short-term memory speech enhancement network (LSTMSE-Net) that exemplifies a sophisticated approach to enhancing speech signals through the integration of audio and visual information. LSTMSE-Net employs a dual-pronged feature extraction strategy, visual features are extracted using a VisualFeatNet comprising a $3$D convolutional frontend and a ResNet trunk  \cite{wang2022retracted}, while audio features are processed using an audio encoder and audio decoder. A key innovation to the system is the fusion of these features to form a comprehensive representation. Visual features are interpolated using bi-linear methods to align with temporal dimensions in the audio domain. This fusion process, combined with advanced processing through a separator network featuring bi-directional LSTMs  
 \cite{hochreiter1997long,graves2005framewise}, underscores the model's capability to effectively enhance speech quality through comprehensive multi-modal integration. The study thus explores new frontiers in AVSE research, aiming to improve intelligibility and fidelity in challenging audio environments.

The evaluation metrics for the model include perceptual evaluation of speech quality (PESQ), short-time objective intelligibility (STOI), and scale-invariant signal-to-distortion ratio (SISDR), with model parameters totalling around $5.1M$which is significantly less than the baseline model (COG-MHEAR Challenge $2024$) which is around $75M$ parameters. The initial model weights are randomized, and the average inference time is approximately $0.3$ seconds per video.

In summary, we have developed a strong AVSE model, LSTMSE-Net, by employing deep learning modules such as neural networks, LSTMs, and convolutional neural netowrks (CNNs). When trained on the challenge dataset provided by the COG-MHEAR challenge 2024, our model obtains higher outcomes across all evaluation metrics despite being substantially smaller than the baseline model provided by the COG-MHEAR challenge 2024. Its decreased size also results in a shorter inference time when compared to the baseline model which takes approximately $0.95$ seconds per video.

\section{Methodology}
\subsection{Overview}
\begin{figure*}
\begin{minipage}{.69\linewidth}
\centering
\subfloat[]{\includegraphics[scale=0.15]{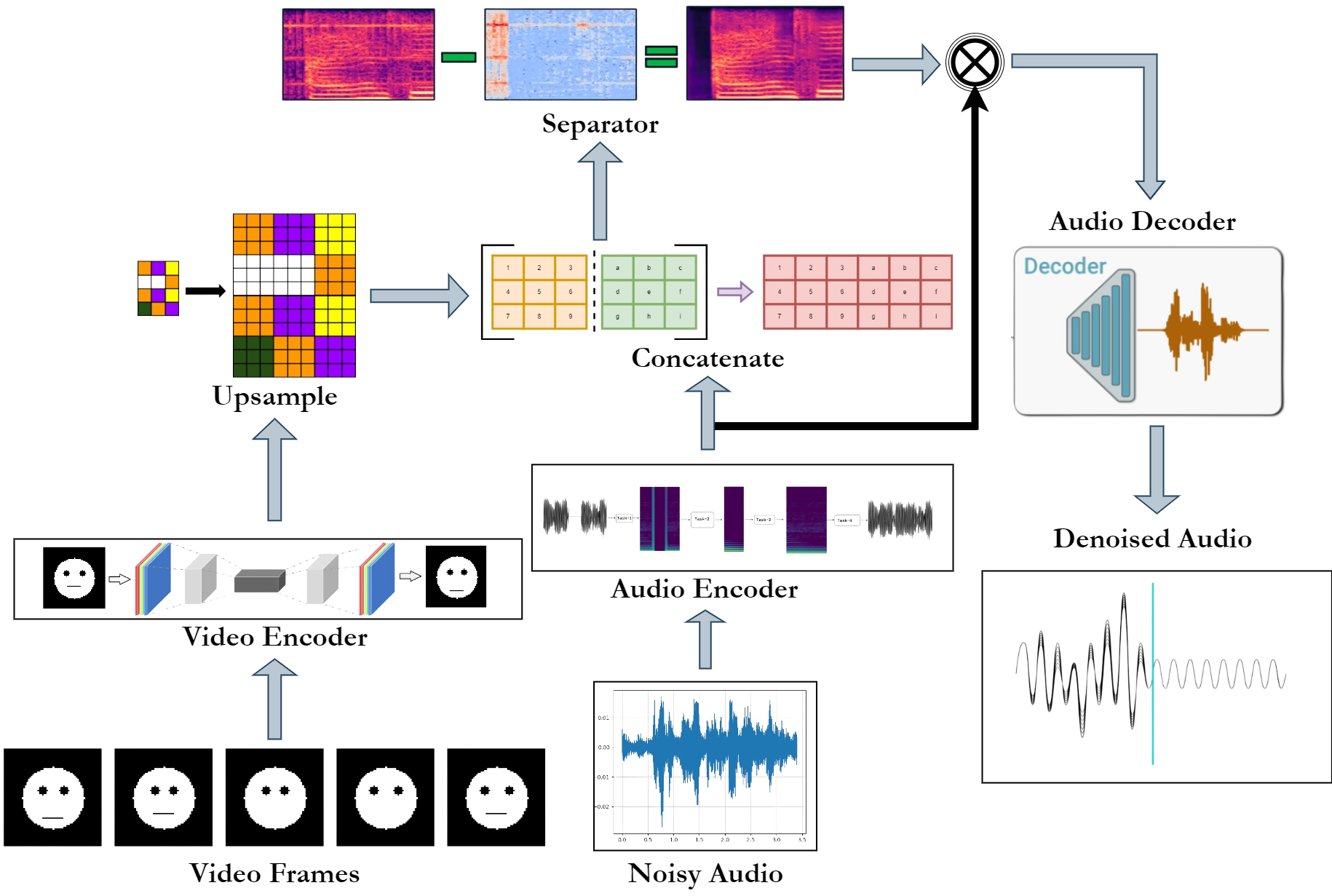}}
\end{minipage}
\begin{minipage}{.29\linewidth}
\centering
\subfloat[]{\includegraphics[scale=0.18]{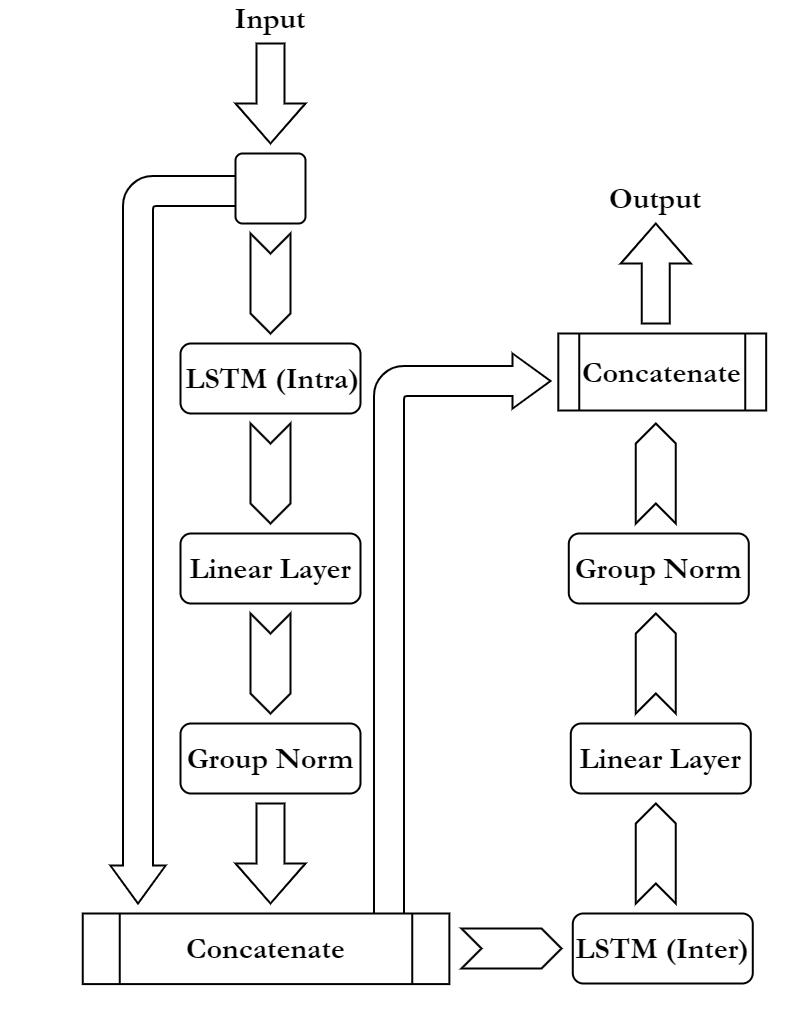}}
\end{minipage}
% \par\medskip
% \par\medskip
\caption{(a) The workflow of the proposed LSTMSE-Net, (b) a single unit in the Separator block of the proposed LSTMSE-Net.
}
\label{fig:LSTM}
\end{figure*}

This section delves into the intricacies of the proposed LSTMSE-Net architecture, which leverages a synergistic fusion of audio and visual features to enhance speech signals. We discuss and highlight its audio and visual feature extraction, integration, and noise separation mechanisms. This is achieved using the following primary components, which are discussed further - audio encoder, visual feature network (VFN), noise separator and audio decoder. The overall architecture of our LSTMSE-Net is depicted in Fig. \ref{fig:LSTM}(a).
% \begin{figure*}[htbp]
%     \centering
%     \includegraphics[width=0.8\textwidth]{AVSE.png}
%     \caption{LSTMSE-Net Workflow}
%     \label{fig:Workflow}
% \end{figure*}
%%%%%%%%%%%%%%%
%%%%%%%%%%%%%%%%%%%%
\subsection{Audio Encoder}
An essential part of the AVSE system, the audio encoder module is in charge of gathering and evaluating audio features. The conv-1d architecture used in this module consists of a single convolutional layer with $256$ output channels, a kernel size of $16$, and a stride of $8$. Robust audio features can be extracted using this setup. To add non-linearity, a rectified linear unit (ReLU) activation function is applied after the convolution step. Afterwards, the upsampled visual features and the encoded audio information are combined and passed into the noise separator.

\subsection{Visual Feature Network}
The VFN is a vital component of the AVSE system, tasked with extracting relevant visual features from input video frames. The VFN architecture comprises a frontend $3$-dimesional ($3$D) convolutional layer, ResNet trunk \cite{he2016deep} and fully connected layers. The 3D convolution layer processes the raw video frames, extracting relevant anatomical and visual features. The ResNet trunk comprises a series of residual blocks designed to capture spatial and temporal features from the video input. The extracted features' dimensionality is then reduced to $256$ by adding a Fully Connected Layer, which simultaneously improves computational complexity and gets the visual features ready for integration with the audio features. 

Bi-linear interpolation is used to upsample the encoded visual characteristics so they match the encoded audio features' temporal dimension. This is done to ensure proper synchronization of features from both modalities. Further, these are concatenated, as mentioned above, with the audio features to form a joint audio-visual feature representation. This is then passed through the separator to extract the relevant part of the audio signal.

\subsection{Feature Extractor and Noise Separator}

\subsubsection{Overview and Motivation}
The Separator module is a crucial component of the LSTMSE-Net, tasked with effectively integrating and processing the combined audio and visual features to isolate and enhance the speech signal. This module leverages Long Short Term Memory (LSTM) networks to capture temporal dependencies and relationships between the audio and visual inputs. The use of LSTM in the AVSE system is further motivated by the following reasons.

Sequential data: Audio and visual features are sequential in nature, with each frame or time step building upon the previous one. LSTM is well-suited to handle such sequential data. Speech signals exhibit long-term dependencies, with phonetic and contextual information spanning multiple time steps. LSTM’s ability to learn long-term dependencies enables it to capture these relationships effectively.

Contextual information: LSTM’s internal memory mechanism allows it to retain contextual information, enabling the system to make informed decisions about speech enhancement.
% \subsubsection{Jasmer here}
\subsubsection{Core functionality and Multimodal ability}
The functionality of the Separator block is based on a multi-modal fusion design. Through the integration of audio and visual inputs, the Separator Block optimizes speech enhancement by utilizing complimentary information from both modalities. The VFN records visual cues including lip movements, which offer important context for differentiating speech from background noise. The ability to identify the portion of audio that the speaker is saying is made easier by the temporal alignment of the visual and aural elements. 

The separator block is made up of several separate units, each of which makes use of intra- and inter-LSTM layers, linear layers, and group normalization. We now elaborate on the information flow in a single unit.
Group normalization layers are used to normalize the combined features following the first feature extraction and concatenation. These normalization steps stabilize the learning process and provide consistent feature scaling, guaranteeing that auditory and visual input are initially given equal priority. The model can recognize complex correlations and patterns between the auditory and visual inputs due to the intra- and inter-LSTM layers. The inter-LSTM layers are intended for a global context, whilst the intra-LSTM layers concentrate on local feature extraction. Through residual connections, the original inputs are brought back to the output of these LSTM layers, aiding in the gradient flow during training and helping to preserve relevant features. More reliable speech enhancement results from the Separator Block's ability to learn the additive and interactive impacts of the audio-visual elements because of this residual design. Fig. \ref{fig:LSTM}(b) shows a single unit of the Separator block. 

As highlighted above the proposed AVSE system employs a multi-modal fusion strategy, combining the strengths of both audio and visual modalities. The final output of the separator module is the mask, which contains only the relevant part of the original input audio features and removes all the background noise. The original input audio features are then multiplied by this mask in order to extract those that are relevant and suppress the ones that are not needed. This generates a clean and processed audio feature map.

\subsection{Audio Decoder}
The audio decoder, which is built upon the ConvTranspose1d \cite{shipton2021implementing} architecture, consists of a single transposed convolution layer with a kernel size of $16$, stride of $8$, and a single output channel. This design facilitates the transformation of the encoded audio feature map back into an enhanced audio signal. It is given the enhanced feature map as the input, and returns the enhanced audio signal, which is also the final model output.

\section {Experiments}
In this section, we begin with a detailed description of the dataset. Next, we outline the experimental setup and the evaluation metrics used. Finally, we present and discuss the experimental results.
\subsection{Dataset Description} 
The data used for training, testing and validation consists of films extracted from the LRS3 dataset \cite{afouras2018lrs3}.
It contains $34524$ scenes (a total of $113$ hours and $17$ minutes) from $605$ speakers appointed for TED and TEDx talks. For the noise, speech interferers were selected from a pool of 405 contestant speakers and $7346$ noise recordings across 15 different divisions.

The videos contained in the test set differ from those used in the training and validation datasets. The train set contains around $5090$ videos or $51$k unique words in vocabulary, whilst the validation and test sets have $4004$ films or $17$k words in its vocabulary and $412$ videos, respectively. 

The dataset has two types of interferers: speech of competing speakers, which are taken from the \textbf{LRS3} dataset (competing speakers and target speakers does not have any overlap) and noise, which is derived from various datasets such as \textbf{CEC1} \cite{graetzer2021clarity}, which consists around $7$ hours of noise, the \textbf{DEMAND} \cite{thiemann2013diverse}, noise dataset includes multi-channel recordings of $18$ soundscapes lasting more than $1$ hour, \textbf{MedleyDB} dataset \cite{bittner2014medleydb} comprises $122$ songs that are royalty-free. Additionally, \textbf{Deep Noise Supression challenge (DNS)} dataset \cite{reddy2020interspeech}, which was released in the previous version of the challenge, features sounds present in AudioSet, Freesound and DEMAND, and {\textbf{Environmental sound classification (ESC-50)} dataset \cite{piczak2015esc} which comprises 50 noise groups that fall into five categories: sounds of animals, landscapes and water, human non-verbal sounds, noises from the outside and within the home, and noises from cities.  Additionally, data preparation scripts are given to us. The output of these scripts consists of the following: S$00001$\_target.wav (target audio), S$00001$\_silent.mp$4$ (video without audio), S$00001$\_interferer.wav (interferer audio), and S$00001$\_interferer.wav (the audio interferer). 

\subsection{Experimental Setup} 
% Our experiments utilized an extensive dataset that includes over 34,000 scenes with 605 target speakers and a wide variety of noise types. The dataset is carefully curated to ensure no overlap between the training and test sets, thereby providing a robust evaluation framework. We employed state-of-the-art hardware for training, ensuring that our model was optimized for performance and efficiency.

We set up our training environment to get the best possible performance and use of the available resources. A rigorous training method that lasted $48$ epochs and $211435$ steps was applied to the model. By utilising GPU acceleration, each epoch took about twenty-two minutes to finish. This effective training length demonstrates how quickly and efficiently the model can handle big datasets.

With $146$ GB of shared RAM, a single NVIDIA RTX A4500 GPU is used for all training and inference tasks. Our LSTMSE-Net model is effectively trained because of it's sturdy training configuration, which also guaranteed that the model could withstand the high computational demands necessary for high-quality audio-visual speech enhancement.

\subsection{Evaluation Metrics}
The LSTMSE-Net model was subjected to a thorough evaluation using multiple standard metrics, including scale-invariant signal-to-distortion ratio (SISDR), short-time objective intelligibility (STOI), and perceptual evaluation of speech quality (PESQ). A comprehensive and multifaceted evaluation is ensured by the distinct insights that each of these measures offers into various aspects of the quality of voice enhancement.

\subsubsection{PESQ}
A standardised metric called PESQ compares the enhanced speech signal to a clean reference signal in order to evaluate the quality of the speech. With values ranging between $-0.5$ to $4.5$, larger scores denote better perceptual quality.

\subsubsection{STOI}
STOI (short-time objective intelligibility) is a metric used to assess how clear and understandable speech is, particularly in environments with background noise. It measures the similarity between the clean and improved speech signals' temporal envelopes, producing a score between $0$ and $1$. Improved comprehensibility is correlated with higher scores.

\subsubsection{SISDR}
By calculating the amount of distortion brought about by the enhancement process, SISDR is a commonly used metric to assess the quality of speech enhancement. Higher SISDR values are indicative of less distortion and improved speech signal quality, making them a crucial indicator for assessing how well our model performs in maintaining the original speech features.

We guarantee a thorough and comprehensive examination of the LSTMSE-Net model by utilising these three complimentary assessment metrics. STOI gauges intelligibility, PESQ assesses perceptual quality, and SISDR concentrates on distortion and fidelity. When taken as a whole, these measurements offer a thorough insight of the model's performance, showcasing its advantages and pinpointing areas that might use improvement. Our dedication to creating a high-performance speech enhancement system that excels in a number of crucial areas of audio quality is demonstrated by this multifaceted evaluation technique.

\subsection{Evaluation Results}
Three types of speeches were included in our evaluation. First, we used the noisy speech provided in the challenge testing dataset, which also served as the audio requiring further enhancement using various AVSE models. Second, we generate the improved speech by applying our LSTMSE-Net model to enhance the noisy speech. Finally, we produce the improved speech using the COG-MHEAR AVSE Challenge 2024 baseline model to enhance the same noisy speech. We evaluated them using the PESQ, STOI, and SISDR, which are the standard evaluation metrics. The table \ref{Tab:Table1} displays the final scores of the models on the evaluation metrics. Compared to noisy speech, the AVSE baseline model produced notably better quality (PESQ) and higher intelligibility (STOI). Furthermore, in PESQ, STOI, and SISDR, LSTMSE-Net outperformed the baseline model by a margin of $0.06$, $0.03$, and $1.32$, respectively. All evaluation criteria showed that LSMTSE-Net performed better than the baseline as well as the noisy speech, which is strong proof of the efficacy of our model. 

The table \ref{Tab:Table2} displays the final inference time of the models on the testing dataset. Compared to the baseline model, which takes an average of $0.95$ secs per video to enhance the audio, LSTMSE-Net only takes $0.3$ seconds per video on average to enhance the audio in them. This significant reduction in processing time underscores the efficiency of LSTMSE-Net.

\begin{table}
\centering
\caption{Comparison of noisy speech, the baseline speech, and LSTMSE-Net (ours) based on PESQ, STOI, and SISDR metrics.}
\resizebox{1\linewidth}{!}{
\begin{tabular}{lccc}
\toprule 
Audio & PESQ & STOI & SISDR \\
\midrule 
Noisy Speech & 1.467288 & 0.610359 & -5.494292 \\
Baseline & 1.492356 & 0.616006 & -1.204192 \\ 
LSTMSE-Net (Ours) & $\mathbf{1.547272}$ & $\mathbf{0.647083}$ & $\mathbf{0.124061}$ \\
\bottomrule
\multicolumn{4}{l}{The boldface in each column denotes the performance of the best}\\
\multicolumn{4}{l}{model corresponding to each metric.}
\end{tabular}}
\label{Tab:Table1}
\end{table}

\begin{table}
\centering
\caption{Comparison between the inference time of the baseline and LSTMSE-Net (ours).}
\begin{tabular}{lccc}
\toprule 
Model & Average inference time per video \\
\midrule 
Baseline & ${0.95}$ seconds \\
LSTMSE-Net (Ours) & $\mathbf{0.3}$ seconds \\
\bottomrule
\end{tabular}
\label{Tab:Table2}
\end{table}

The superior efficiency and efficacy of the proposed LSTMSE-Net not only reduces the computational load but also enables real-time processing, making it highly suitable for applications requiring low latency. Moreover, the smaller model size enhances scalability, allowing the deployment of LSTMSE-Net on a wider range of devices, including those with limited computational resources. This makes the proposed model an excellent choice for both high-performance systems and resource-constrained environments, demonstrating its versatility and practical applicability.
%%%%%%%%%%%%%%%%%%%%%%
\section{Conclusion and Future Work}
This research presents LSTMSE-Net, an advanced AVSE architecture that improves speech quality by fusing audio signals with visual information from lip movements. The LSTMSE-Net architecture consists of an audio decoder, a visual encoder, a separator, and an audio encoder. Each of these components is essential to the processing and refinement of the input signals in order to generate enhanced speech that is high-quality.
% The audio encoder efficiently captures salient features from the noisy audio input, while the visual encoder, based on a modified ResNet structure, extracts complementary features from the video frames. These features are concatenated and processed by the separator module, which utilizes intra- and inter-sequence LSTM layers to refine the integrated features and effectively isolate the clean speech signal. The audio decoder then reconstructs the enhanced speech waveform from the refined features.
AVSE-Net exhibits the capacity to efficiently capture and leverage both local and global audio-visual interdependence. With the use of advanced deep learning methods like convolutions and long short term memory networks, LSTMSE-Net improves voice enhancement significantly.

Experimental studies on the benchmark dataset, i.e., COG-MHEAR LRS3 dataset, confirm LSTMSE-Net's superior performance. LSTMSE-Net performs much better than baseline models on the COG-MHEAR LRS3 dataset, demonstrating its effectiveness in combining visual and aural characteristics for improved speech quality. To sum up, LSTMSE-Net is a major breakthrough in audio-visual speech improvement, utilising the complementary qualities of both auditory and visual input to provide better voice quality. This work establishes a new benchmark in the field by offering a scalable and efficient solution for speech improvement.

\noindent For our future work, we have the following plans:
\begin{itemize}
    \item We aim to extend our model to incorporate causality in its architecture, enabling real-time deployment. This enhancement will ensure that the model relies solely on past and current information for predictions.
    \item We plan to propose an enhanced version of LSTMSE-Net that incorporates attention mechanisms and advanced feature fusion techniques to further refine the integration of visual and audio features. Our goal is to achieve superior performance across various AVSE benchmarks.
    \item Additionally, we will conduct a comprehensive comparative analysis of LSTMSE-Net and other state-of-the-art AVSE variants. This analysis will focus on their performance in real-world noisy environments to identify strengths and areas for improvement.
\end{itemize}

\section{Acknowledgement} The authors are grateful to the anonymous reviewers for their invaluable comments and suggestions. This project is supported by the Indian government’s Science and Engineering Research Board (SERB) through the Mathematical Research Impact-Centric Support (MATRICS) scheme under grant MTR/2021/000787. Prof Hussain acknowledges the support of the UK Engineering and Physical Sciences Research Council (EPSRC) Grants Ref. EP/T021063/1 (COG-MHEAR) and EP/T024917/1 (NATGEN). The work
of M. Sajid is supported by the Council of Scientific and Industrial Research (CSIR), New Delhi for providing fellowship under the under Grant
09/1022(13847)/2022-EMR-I.

\bibliographystyle{IEEEtran}
\bibliography{mybib}

\end{document}